\documentclass[%
 reprint,
superscriptaddress,
showpacs,preprintnumbers,
 amsmath,amssymb, 
 aps,
 prd,
 abbrv,
]{revtex4-1}

\usepackage{cancel}
\usepackage{accents}
\usepackage{footmisc}
\usepackage{mciteplus,slashed}
\usepackage{amssymb,cancel,amsmath}
\usepackage{dcolumn}
\usepackage{bm}
\usepackage[caption=false]{subfig}
\usepackage{appendix}
\usepackage{physics}
\usepackage{feynmp-auto}
\usepackage[T1]{fontenc}	
\usepackage{csvsimple}
\usepackage{hyperref}
\usepackage[section]{placeins}
\usepackage[capitalise]{cleveref}
\usepackage{booktabs}
\usepackage{graphicx}
\usepackage{mathrsfs}
\graphicspath{{Figures/}}

\usepackage[dvipsnames]{xcolor}
\usepackage[normalem]{ulem}

\newcommand{\NA}{\text{---}}

\newcommand{\be}{\begin{equation}}
\newcommand{\ee}{\end{equation}}
\newcommand{\ba}{\begin{align*}}
\newcommand{\ea}{\end{align*}}
\newcommand{\bpm}{\begin{pmatrix}}
\newcommand{\epm}{\end{pmatrix}}
\newcommand{\bea}{\begin{eqnarray}}
\newcommand{\eea}{\end{eqnarray}}

\newcommand{\benum}{\begin{enumerate}}
\newcommand{\eenum}{\end{enumerate}}
\newcommand{\bi}{\begin{itemize}}
\newcommand{\ei}{\end{itemize}}

\newcommand{\MeV}{~\mathrm{MeV}}
\newcommand{\GeV}{~\mathrm{GeV}}

\newcommand{\gsim}{\lower.7ex\hbox{$\;\stackrel{\textstyle>}{\sim}\;$}}
\newcommand{\lsim}{\lower.7ex\hbox{$\;\stackrel{\textstyle<}{\sim}\;$}}

\def\Ageo{A_\text{geo}}

\begin{document}

\title{
Millicharged particles in neutrino experiments}

\author{Gabriel Magill}
\email{gmagill@perimeterinstitute.ca}
\affiliation{Perimeter Institute for Theoretical Physics, 31 Caroline Street North, Waterloo, Ontario N2L 2Y5, Canada}
\affiliation{\mbox{Department of Physics \& Astronomy, McMaster University, 1280 Main Street West, Hamilton, Ontario L8S 4M1, Canada}}
\author{Ryan Plestid}
\email{plestird@mcmaster.ca}
\affiliation{Perimeter Institute for Theoretical Physics, 31 Caroline Street North, Waterloo, Ontario N2L 2Y5, Canada}
\affiliation{\mbox{Department of Physics \& Astronomy, McMaster University, 1280 Main Street West, Hamilton, Ontario L8S 4M1, Canada}}
\author{Maxim Pospelov}
\email{mpospelov@perimeterinstitute.ca}
\affiliation{Perimeter Institute for Theoretical Physics, 31 Caroline Street North, Waterloo, Ontario N2L 2Y5, Canada}
\affiliation{Department of Physics and Astronomy, University of Victoria, Victoria, BC V8P 5C2, Canada}
\affiliation{Theoretical Physics Department, CERN, 1211 Geneva, Switzerland}
\author{Yu-Dai Tsai}
\email{ytsai@fnal.gov}
\affiliation{Fermilab, Fermi National Accelerator Laboratory, Batavia, IL 60510, USA}

\date{\today}

\begin{abstract}
We set constraints and future sensitivity projections on millicharged particles (MCPs) based on electron scattering data in numerous neutrino experiments, starting with MiniBooNE and the Liquid Scintillator Neutrino Detector (LSND). Both experiments are found to provide new (and leading) constraints in certain MCP mass windows: $5\:-\:35\MeV$ for LSND and $100\:-\:180\MeV$ for MiniBooNE. Furthermore, we provide projections for the ongoing Fermilab SBN program, the Deep Underground Neutrino Experiment (DUNE), and the proposed Search for Hidden Particles (SHiP) experiment. In the SBN program, SBND and MicroBooNE have the capacity to provide the leading bounds in the $100\:-\:300\MeV$ mass regime. DUNE and SHiP are capable of probing parameter space for MCP masses in the range of $5\MeV-5\GeV$ that is significantly beyond the reach of existing bounds, including those from collider searches and, in the case of DUNE, the SLAC mQ experiment. 
\end{abstract}
 
\maketitle

\section{Introduction} 
Extensions of the Standard Model (SM) by weakly interacting particles, and their probes at the intensity frontier experiments have become an important direction of particle physics \cite{Battaglieri:2017aum}. One of the simplest and most natural ways of coupling new particles to the SM is via a ``kinetic mixing'' or ``hypercharge portal'' \cite{Holdom:1985ag,Izaguirre:2015eya}, which at low energies may lead to millicharged particles (MCPs), that would seemingly contradict the observed quantization of electric charge in nature \cite{Dobroliubov:1989mr}. In recent years, a wide class of related models were studied in connection with dark matter \cite{Brahm:1989jh,Boehm:2003hm,Pospelov:2007mp} (see also \cite{Bjorken2009,Batell2009,deNiverville:2011it,Izaguirre:2013uxa,Batell:2014mga,Kahn:2014sra,Dobrescu:2014ita,Coloma:2015pih,deNiverville:2016rqh,Ge:2017mcq}), and MCPs can be viewed as a specific limit of those theories. 

It is well appreciated that both proton and electron beam dump experiments provide sensitive probes of vector portal models.  
 In particular, production and scattering of light dark matter \cite{Batell2009} has been studied as a function of mediator mass $m_{A'}$, dark sector coupling $\alpha_D$, dark matter mass $m_\chi$, and kinetic mixing parameter $\epsilon_Y$. Depending on the relation between these parameters, either the past electron beam dump facilities \cite{Batell:2014mga} or the proton fixed-target experiments with a primary goal of neutrino physics \cite{deNiverville:2011it,Kahn:2014sra,Arguelles:2019xgp} provide the best sensitivity. The simplest limit of $m_{A'}\to 0$, when the parameter space simplifies to the mass and effective charge of MCPs, $\{m_\chi, \epsilon \}$, was analyzed only in the context of electron fixed-target experiments \cite{Prinz:1998ua,Prinz:2001qz,Davidson:2000hf}, despite earlier studies of proton fixed-target experiments' potential as a probe of MCPs \cite{Golowich1987,Babu1993,Gninenko2006}. Clearly, fixed target neutrino experiments, such as the existing data from MiniBooNE \cite{Aguilar-Arevalo:2018gpe} and the Liquid Scintillator Neutrino Detector (LSND) \cite{Athanassopoulos:1996ds}, and the soon to be released data from MicroBooNE, the ongoing SBN program \cite{Antonello:2015lea}, the Deep Underground Neutrino Experiment (DUNE) \cite{DUNECollaboration2015}, and the proposed Search for Hidden Particles (SHiP) \cite {Anelli:2015pba} serve as a fertile testing ground of MeV--GeV physics due to their high statistics \cite{deNiverville:2011it, Kahn:2014sra, Pospelov:2017kep, Magill:2018jla, Arguelles:2018mtc,Tsai:2019mtm}.  These experiments all serve as promising avenues to probe MCPs. 

The purpose of this work is twofold: First, we demonstrate that existing data from LSND provides leading bounds on MCPs (surpassing existing constraints from SLAC's mQ experiment \cite{Prinz:1998ua}) in the low mass regime ($m_\chi\lesssim 35~\text{MeV}$). Likewise, newly released data from MiniBooNE \cite{Aguilar-Arevalo:2018gpe} represents the current leading bounds on MCPs in the mass range of $100\MeV \lesssim m_\chi \lesssim 180\MeV$ (offering better sensitivity than collider experiments \cite{Davidson:2000hf,Haas:2014dda}). 
Second, we predict that by optimizing search strategies at ongoing and upcoming experiments (such as MicroBooNE, SBND, DUNE, and SHiP), fixed source neutrino experiments can serve to provide leading bounds for MCPs over the full range of masses $5~\text{MeV}\lesssim m_\chi\lesssim 5~\text{GeV}$. The detection signature of MCPs in these experiments is elastic scattering with electrons, and we find that detection prospects are highly sensitive to the threshold imposed on the electron's recoil energy. Therefore, significant gains in sensitivity to MCPs may be achieved by future experiments by optimizing the detection of low-energy electrons. 
Our setup can be easily extended to probe light dark matter coupling to a massive dark photon.
\begin{figure}[!t]
\includegraphics[width=\linewidth]{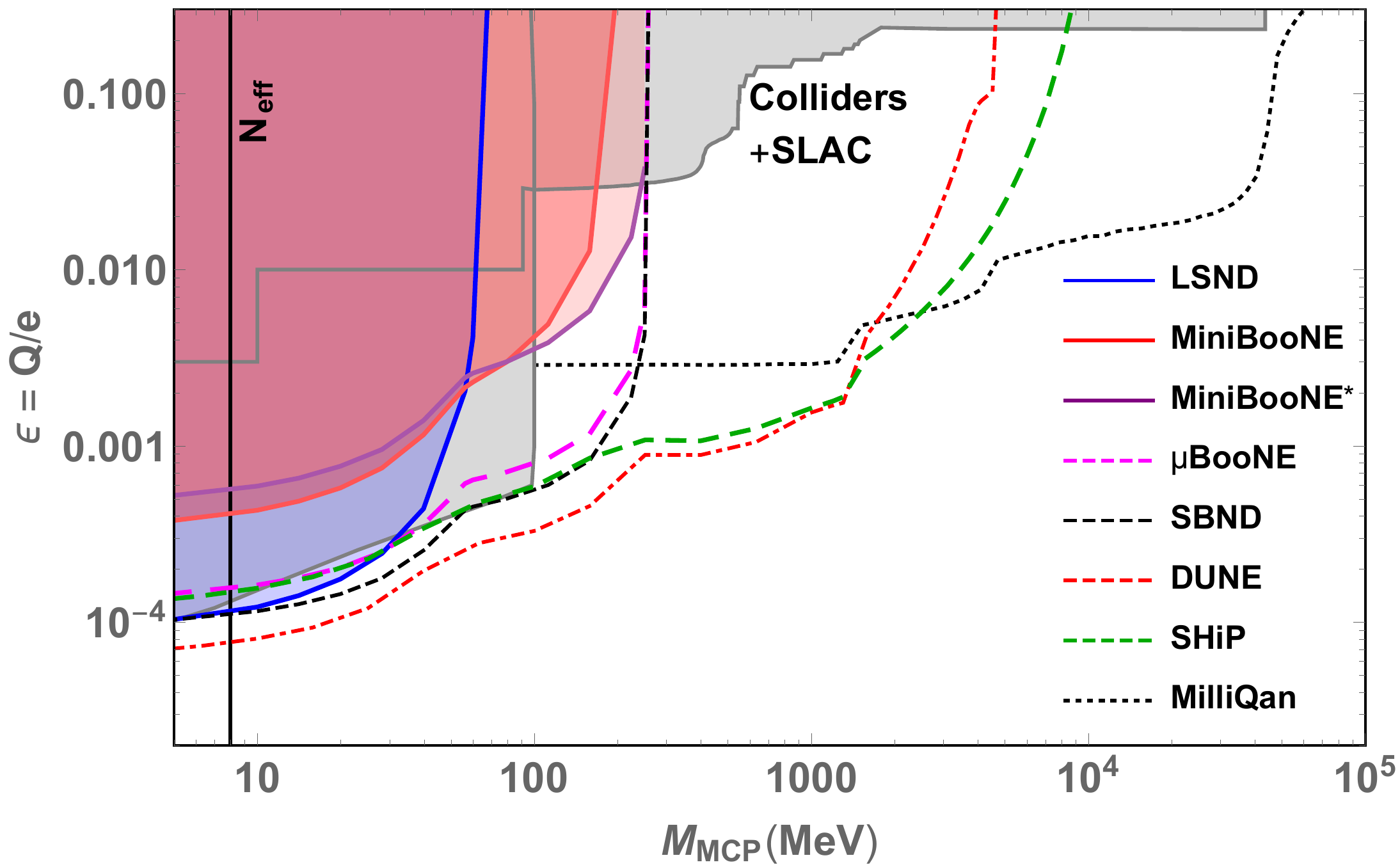}
\caption{Exclusion curves for fermionic MCPs (results are broadly similar for scalars). Existing data is shown as solid lines, while projections are shown as dashed curves. 
The kinematic reach of a given experiment is set by the heaviest meson of interest it can produce. This is $\pi^0$ for LSND, $\eta$ for the Booster experiments, and $\Upsilon$ for DUNE. At SHiP, Drell-Yan production extends the kinematic reach to roughly $10\GeV$. The sensitivity of each experiment can be understood via \cref{eq:NumberSignalEvents} while the relevant parameters for each experiment are summarized in \cref{tab:summary}. 
The bound on $N_\text{eff}$ \cite{Boehm:2013jpa} comes from changing the effective number of neutrinos, while the SLAC bound (below $\sim$ 100 MeV) and the collider bound (extends up to $\sim$ 30 GeV in the plot) are taken from \cite{Prinz:1998ua} and \cite{Davidson:2000hf,Haas:2014dda} respectively. The projected sensitivities at milliQan are from \cite{Ball:2016zrp,Haas:2014dda}. Our exclusions apply independently of the existence of a dark photon, which introduces additional constraints \cite{Berlin:2018sjs}.} 
\label{fig:masterPlot}
\end{figure}

Our results have direct implications for models with late kinetic coupling of dark matter and baryons \cite{Tashiro:2014tsa} that could lead to extra cooling of the baryon fluid and spin temperature at redshifts $z\sim 20$, which may result in a more pronounced 21\,cm absorption signal. If a fraction of dark matter is composed of MCPs, this extra cooling mechanism can be naturally realized \cite{Barkana:2018lgd,Munoz2018}, and fit the unexpected signal reported by Experiment to Detect the Global Epoch of Reionization Signature (EDGES) \cite{Bowman:2018yin}. 
Our analysis reveals that sensitivities from LSND, SBND, SHiP, and DUNE can explore previously unprobed regions of parameter space that are favored by the $1\%$-MCP fractional dark matter hypothesis \cite{Munoz2018,Berlin:2018sjs,Barkana:2018qrx}, as well as, a proposal to use MCPs to reduce electron number (while maintaining charge neutrality) to achieve an earlier decoupling of the baryon gas from the CMB \cite{Falkowski:2018qdj}. 


\section{Production and detection}
Fixed-target neutrino experiments rely on the production of neutrinos from weak decays of charged pions. In generating a large flux of $\pi^\pm$ these experiments necessarily also produce a similar number [i.e.\ $\mathcal{O}(10^{20})$] of $\pi^0$ \cite{deNiverville:2016rqh}. For large beam energies, other neutral mesons (e.g.\ $\eta$, $\Upsilon$, $J/\psi$) are also produced. Significant branching ratios to lepton pairs necessarily imply associated decays to pairs of MCPs, resulting in a significant flux of MCPs even for extremely small charges. In the case of $\eta$ and $\pi^0$, Dalitz decays $\pi^0,~\eta\rightarrow \gamma \chi \bar{\chi}$ dominate, while for $J/\psi$ and $\Upsilon$ direct decays, $J/\psi,\Upsilon\rightarrow \chi\bar{\chi}$ are the most important. The branching ratio for a meson, $\mathcal{M}$, to MCPs is given roughly by
\begin{equation}
\text{BR}\qty(\mathcal{M}\rightarrow \chi\bar{\chi})\approx \epsilon^2 \times \text{BR}\qty(\mathcal{M}\rightarrow X e^+e^-)\times f\qty(\frac{m_\chi}{M}),
\label{eq:branching}
\end{equation} 
where $M$ is the mass of the parent meson, $X$ denotes any additional particles, and $f(m_\chi/M)$ is a phase space factor that decreases slowly as a function of $m_\chi/M$. For the 2-body decay cases, $f(m_\chi/M)$ has an simple analytic form $\sqrt{1-\frac{m^2_\chi}{M^2}}$. The number of MCPs passing through the detector is a function of both the branching ratio and geometric losses which varies substantially between experiments (see \cref{tab:summary}).

We now turn to detecting MCPs in neutrino experiments, where the predominant signature is elastic scattering with electrons. Electron scattering dominates the detection signal because of the low-$Q^2$ sensitivity of the scattering cross section. 
Explicitly, in the limit of small electron mass, we have
\begin{equation}
\dv{\sigma_{e\chi}}{Q^2}=2 \pi \alpha ^2 \epsilon^2 \times \frac{ 2 (s-m_\chi^2)^2 -2 s Q^2+Q^4}{(s-m_\chi^2)^2 Q^4}.
\label{eq:dsigmadt}
\end{equation}
Integrating over momentum transfers, we see that the cross section is dominated by the small-$Q^2$ contribution to the integral. In this limit, we have $\dd\sigma_{e\chi}/\dd Q^2\approx 4\pi \alpha^2 \epsilon^2/Q^4$, and therefore $\sigma_{e\chi}\approx4\pi \alpha^2 \epsilon^2/Q^2_\text{min}$. In the lab frame $Q_\text{min}$ is related to the recoil energy of the electron via $Q^2=2m_e( E_e -m_e)$ \footnote{Note that for nucleon scattering, the cross section $\sigma \propto 1/Q^2$ is suppressed by $1/m_p$ rather than $1/m_e$.}.
An experiment's recoil energy threshold, $E_e^{(\text{min})}$, then sets the scale of the detection cross section as
\begin{equation}
\sigma_{e\chi} \approx 2.6\times 10^{-25}\text{cm}^2\times\epsilon^2 \times \frac{1\,\MeV}{E^{(\text{min})}_e-m_e}.
\label{eq:cross-section}
\end{equation}
Consequently, sensitivity to MCPs can be enhanced by  measuring low electron energy recoils (an important feature for future search strategies).

\section{Results}
The various curves  in  \cref{fig:masterPlot} are obtained by performing a sensitivity analysis \cite{Amsler:2008zzb}: given a number of predicted background events $b$ and data $n$, the number of signal events $s_\text{up}$ consistent with the observation and backgrounds at $(1-\alpha)$ credibility level is found by solving the equation $\alpha = \Gamma(1+n,b+s_\text{up})/\Gamma(1+n,b)$ where $\Gamma(x,y)$ is the upper incomplete gamma function \cite{NoteX}. 
%
Using a credibility interval of $1-\alpha=95\%$ we calculate the corresponding bounds implied by $s_\text{up}$ according to the formula
\begin{equation}
s_\text{up}=\sum_\text{Energies}
N_\chi(E_i) \times \frac{N_e}{\text{Area}}\times \sigma_{e\chi}(E_i;~m_\chi)\times\mathcal{E}.
\label{eq:NumberSignalEvents}
\end{equation}
\Cref{eq:branching,eq:cross-section} implies that $s_\text{up}\propto \epsilon^4$. Here, $\epsilon$ is the MCP electric charge (in units of $e$), $N_\chi(E_i)$ is the number of MCPs with energy $E_i$ that pass through the detector, $\sigma_{e\chi}(E_i)$ is the detection cross section (with the angular and recoil cuts imposed), $N_e$ is the number of electrons inside the detector's active volume, $\mathcal{E}$ is the electron detection efficiency. ``Area'' in (\ref{eq:NumberSignalEvents}) stands for the active volume divided by the average length $\langle { l} \rangle$ traversed by particles inside the detector. 
\begin{table}
\centering
$\begin{array}{ cccccccc }
\multicolumn{1}{c}{~} &
\multicolumn{2}{c}{\text{$N$ [$\times 10^{20}$]} } &
\multicolumn{2}{c}{\Ageo(m_\chi)[\times10^{-3}] } &
\multicolumn{2}{c}{\text{Cuts [MeV]}} \\ 
\cmidrule(lr){2-3}
\cmidrule(lr){4-5}
\cmidrule(lr){6-7}
\text{Exp.} 		&~\pi^0	& \eta	& 1\MeV 	& 100\MeV 	& E^{\text{min}}_e	& E^{\text{max}}_e 	& \text{Bkg} 		\\
\midrule
\text{LSND}		& 130 	        & \NA	& 20 		& \NA 		& 18 				& 52 	        & 300	                \\
\text{mBooNE}		& 17		& 0.56	& 1.2		& 0.68		& 130 			        & 530		& 2\text{k} 		\\
\text{mBooNE*}		& 1.3		& 0.04	& 1.2		& 0.68		& 75	& 850 		& .4^{*} 	 		\\
\mu\text{BooNE} 	& 9.2	 	& 0.31	& 0.09	        & 0.05		& 2				& 40 	 	& 16           	 	\\
\text{SBND}		& 4.6		& 0.15	& 4.6		& 2.6	 	& 2 				& 40 	        & 230 			\\
\text{DUNE}		& 830	        & 16	& 3.3		& 5.1		& 2 				& 40 		& 19\text{k}		\\
\text{SHiP}		& 4.7		& 0.11 	& 130	        & 220		& 100			        & 300    	& 140 			\\
 \bottomrule
\end{array}$
\caption{Summary of the lifetime meson rates ($N$), MCP detector acceptances ($\Ageo$), electron recoil energy cuts, and backgrounds at each of the experiments considered in this paper. In all experiments a cut of $\cos\theta>0$ is imposed in our analysis ($^*$except for MiniBooNE's dark matter run where a cut of $\cos\theta>0.99$ reduces the SM background to 0.4 \cite{Aguilar-Arevalo:2018wea,Dharmapalan:2012xp}). For the SHiP and DUNE experiments, we also include $J/\psi$ and $\Upsilon$ mesons as well as Drell-Yan production which are discussed in the text. We use an efficiency of $\mathcal{E}=0.2$ for Cherenkov detectors, $\mathcal{E}=0.5$ for nuclear emulsion detectors, and $\mathcal{E}=0.8$ for liquid argon time projection chambers. The data at LSND and MiniBooNE is taken from \cite{Auerbach:2001wg} and \cite{Aguilar-Arevalo:2018wea,Aguilar-Arevalo:2018gpe} respectively. Projections at
MicroBooNE \cite{Acciarri:2017sjy}, SBND \cite{Antonello:2015lea}, DUNE \cite{DUNECollaboration2015} and SHiP \cite{Alekhin:2015byh} are based on expected detector performance.}
\label{tab:summary}
\end{table}
Electromagnetic decays of mesons dominate the flux for $m_\chi\lesssim m_\eta/2$, whereas Drell-Yan production (DYP) dominates for  large MCP masses (only accessible at DUNE and SHiP). 

To estimate how many MCPs of energy $E_i$ arrive at the detector, we model the angular and energy distributions of the mesons using empirical formulas (discussed below) that parameterize the double-differential distribution of mesons (e.g.\ $\dd^2 N_\pi/\dd\Omega \dd E_\pi$) in the lab frame. Then,  for each meson produced at a given angle and energy, we calculate the differential branching ratio with respoect to the energy and angle of the MCP in the lab frame.  This generates a double-differential distribution of MCPs in the lab frame for each meson. Next, using this new MCP-distribution, we determine the fraction of MCPs which pass through the detector with an energy within a bin centered at $E_i$ giving us a histogram-spectrum of MCPs as a function of $E_\chi$.  Repeating this procedure over all production energies and angles of the meson, and appropriately weighting the contribution by the meson-distribution, yields each meson's contribution to $N_\chi(E_i)$. DYP of MCPs from a quark and anti-quark pair is treated similarly, but to generate the MCP-distribution we integrate over the full production phase-space using MSTW parton distribution functions \cite{Martin2009}, and using Heaviside functions, we select the proportion of events containing an MCP pointed towards the detector, with energy $E_i$. 

Our main qualitative (but not quantitative) results can be understood without appealing to the details above, by introducing the quantity  $\Ageo(m_\chi)$ which is defined as the ratio between the total number of MCPs that reach the detector and the total number of parent-mesons produced \footnote{The number of parent mesons depends on the MCP mass due to threshold effects (e.g.\ for $m_\chi > m_\pi/2$ pions are no longer included as parent mesons)}. Given the number of parent mesons at a given neutrino experiment and $A_\text{geo}$, then gives a rough estimate of the total number of MCPs that pass through the detector (e.g.\ at LSND for $m_\chi=1\MeV$ we have $\Ageo \approx 2\%$ and $N_{\pi^0}=1.3\times10^{22}$ and so  $N_\chi^{\text{tot}}\approx\sum_i N(E_i)\approx 2.6\times 10^{20}$.
Using  the information in \cref{tab:summary} and \cref{eq:NumberSignalEvents}  one can approximately reproduce our results.


At LSND, the $\pi^0$ spectrum is modeled using a Burman-Smith distribution \cite{Burman_Smith_1989,BURMAN1990621} assuming 2 years of operation on a water target and 3 years of operation on a tungsten target. Our LSND analysis is based on \cite{Auerbach:2001wg}, which featured $1.7\times 10^{23}$ protons on target (POT), a beam energy of 0.798 GeV, and a single electron background of approximately 300 events with energies ranging between $18$ MeV and $52$ MeV. We estimate the $N_e/$Area in \cref{eq:NumberSignalEvents} to be $2.5\times 10^{26}~e^{-}/\text{cm}^2$. 

The meson spectrum from Fermilab's Booster Neutrino Beam (BNB) is relevant for MiniBooNE, MicroBooNE, and SBND. The BNB delivers 8.9 GeV POT and so can produce substantial numbers of both $\pi^0$ and $\eta$ mesons. The former's angular and energy spectra are modeled by the Sanford-Wang distribution \cite{deNiverville:2016rqh,PhysRevD.79.072002}, and $\eta$ mesons by the Feynman Scaling hypothesis \cite{PhysRevD.79.072002}. These distributions are common across all three experiments. Our geometric acceptances are in reasonable $\mathcal{O}(1)$ agreement with \cite{deNiverville:2016rqh}.

At MiniBooNE we perform two analyses: First, we consider the recently updated neutrino oscillation search \cite{Aguilar-Arevalo:2018gpe}. We combine data from both neutrino and anti-neutrino runs and consider a sample of $2.41\times 10^{21}$ POT with a single electron background of $2.0\times 10^3$ events and a measured signal rate of $2.4 \times 10^3$. Next, we consider a parallel analysis involving electron-recoil data with $1.86\times 10^{20}$ POT \cite{Aguilar-Arevalo:2018wea}. Backgrounds were suppressed by operating the beamline in an ``off-target'' (i.e.\ not collimating charged pions) beam-dump mode, and these are further suppressed (to 0.4 predicted events) by imposing a cut of $\cos\theta>0.99$ on the electron's recoil angle \cite{Aguilar-Arevalo:2018wea,Dharmapalan:2012xp}.  In both cases we estimate an electron number density of $3.2 \times 10^{26}~e^-/\text{cm}^2$ and do not include timing cuts in our analysis. 

At MicroBooNE, the meson rates assume $1.32\times 10^{21}$ POT and we estimate an electron density of $3.9 \times 10^{26}~e^-/\text{cm}^2$. The chosen recoil cuts are based on the lowest reaches achievable given the wire spacing in MicroBooNE's liquid argon detector \cite{Acciarri:2017sjy}. The wire spacing is 3 mm and the ionization stopping power is approximately $2.5\MeV/\text{cm}$, so electrons with total energy larger than at least 2 MeV produce tracks long enough to be reconstructed across two wires. Requiring that ionization signals don't shower limits our recoil window to be between $2$ MeV and $40$ MeV. The treatment of SBND is similar, except we assume $6.6\times 10^{20}$ POT, which corresponds to half the run time of MicroBooNE.

At SHiP we assume $2\times 10^{20}$ POT and a near neutrino detector $50$ m from the beam stop with an electron density of $2.7\times10^{26}~e^-/\text{cm}^2$. We include $J/\psi$ and $\Upsilon$, in addition to $\pi^0$ and $\eta$. We do not include mesons such as $\rho$, $\omega$ and $\phi$, because they do not serve to significantly alter the sensitivity offered by $J/\psi$. 
At SHiP's energies, production of $\pi^0$ and $\eta$ is described by the BMPT distributions \cite{deNiverville:2016rqh,BMPT:Bonesini:2001iz}. 
We have compared our geometric acceptances to those obtained using \cite{deNiverville:2016rqh} and found reasonable agreement, with our acceptances being smaller by a factor of four. For production of $J/\psi$, we use the energy production spectra described in \cite{Gale:1998mj}. These distributions rely on production being highly peaked in the forward direction and parameterized as $\dd \sigma/\dd x_F\propto (1-|x_F|)^5$, where $x_F=2p_{\parallel}/\sqrt{s}$ is the meson's longitudinal component in the COM frame of the collision. We account for geometric losses by using an empirical formula for the $p_T$ distribution provided in \cite{Gribushin:1999ha}. We assume that the production spectrum of $\Upsilon$ mesons are similarly given, and normalize their total cross section to the data in \cite{Abt:2006wc}. For $J/\Psi$ we have reproduced the Pb rates in Table 3 of \cite{Alessandro:2006jt}, while for $\Upsilon$ we reproduced the Pt rates in Table 1 of \cite{Herb:1977ek}. We estimate $N_{J/\psi}=2.1\times 10^{15}$ with an acceptance of $\Ageo(100~\text{MeV})=8\times 10^{-2}$, and $N_{\Upsilon}=1.2\times 10^{11}$ with $\Ageo(100~\text{MeV})=7.2\times 10^{-2}$. For large MCP masses, DYP becomes the main production mechanism, and we calibrate our DYP calculations by reproducing the dimuon invariant mass spectrum in Fig.\ 11 of \cite{Stirling:1993gc} from the FNAL-772 experiment \cite{Alde:1990im}.

At DUNE, our treatment of meson production is very similar to the treatment at SHiP. We model pseudoscalar meson production using the BMPT distribution, but use a beam energy of $80$ GeV \cite{DUNECollaboration2015} and account for differences in the target material. We also include $J/\psi$ and $\Upsilon$ mesons and treat them as described above. Our detector treatment and electron recoil cuts are motivated by MicroBooNE's liquid argon time projection chamber (LAr-TPC) detector, and its ability to measure low-energy electron recoils. We assume $3\times 10^{22}~$ POT and a 30 tonne liquid argon detector which corresponds to $5.4\times10^{25}~e^-/\text{cm}^2$. We estimate $N_{J/\psi}=3\times 10^{16}$ with an acceptance of $\Ageo(100~\text{MeV})=2.4\times 10^{-3}$ and $N_{\Upsilon}=5.1\times 10^{9}$ with $\Ageo(100~\text{MeV})=3.7\times 10^{-3}$. It is important to point out that the attenuation of MCPs through dirt and rock only becomes important for MCPs lighter than $\sim$ MeV \cite{Prinz:2001qz} with large $\epsilon$ and is neglected in our analysis. This could have small effects on the MCP fluxes in all mass ranges and deserves a seperate and dedicated study.
We also neglect the multiple scatterings of MCPs inside the detectors, which could be used as an additional tool of discriminating their signature against the neutrino background. 

 We consider two classes of backgrounds for \cref{tab:summary}: those coming from each experiments flux of neutrinos [i.e.\ $\nu e\rightarrow \nu e$ and $\nu n\rightarrow e p$], and those coming from sources such as cosmics, mis-identified particles, or dirt related events. 

We treat neutrino induced backgrounds by summing over the neutrino fluxes from each collaboration and accounting for the detection efficiencies $\mathcal{E}$. A large background reduction is obtained by imposing the electron recoil cuts $E_e^{(\text{max})}$ shown in \cref{tab:summary}. These do not significantly affect the signal (which is dominated by low-energy electron recoils), but \emph{do} significantly reduce charged and neutral current backgrounds \cite{Strumia:2003zx,ccqeccqe}. In the case of MiniBooNE's dark matter run based on electron recoils, the angular requirements already provide a powerful background discriminant and the maximum energy of the electron is determined by the kinematics of the event.

External sources of backgrounds are estimated by multiplying the neutrino-induced backgrounds by an overall multiplicative factor. LAr-TPC detectors can use timing information as vetoes to reduce additional sources of backgrounds; this is not possible in a nuclear emulsion chamber. Therefore, we multiply our neutrino induced backgrounds by a factor of 10 for LAr-TPC detectors (MicroBooNE, SBND, and DUNE) and a factor of 25 for nuclear emulsion detectors (SHiP); this increase in the backgrounds decreases our sensitivity to $\epsilon$ by $20-30\%$. We have chosen an overall multiplicative factor because this naturally scales the backgrounds with the overall size of each detector. We anticipate that the timing information in  LAr-TPC detectors, coupled with their fantastic vertex resolution, will allow it to veto backgrounds more effectively than nuclear emulsion detectors whose tracks do not contain any timing information; 
we have consequently chosen the factors of 10 and 25 respectively. We emphasize that our results in \cref{fig:masterPlot} can be easily revised for different background assumptions according to \cite{NoteX}.

\section{Outlook} 
Fixed-target neutrino experiments can probe MCPs due to the large number of mesons produced with electromagnetic decay pathways. Using existing data, LSND and MiniBooNE are able to provide the leading sensitivity to MCPs for certain sub-GeV masses. Beyond serving as a test of fundamental physics, this newfound sensitivity has implications for models of physics beyond the Standard Model. In particular it further restricts the parameter space of cosmological models where a fraction of MCP dark matter results in extra cooling of baryons that modifies 21\,cm physics at high redshifts. 

Equally important are our projected sensitivities at MicroBooNE, SBND, DUNE, and SHiP. The success of these experiments as probes of MCPs will rely heavily on their respective collaborations' search strategies. Increasing the sensitivity to low-energy electron recoils can be enhance the signal with a scaling proportional to $1/(E_e-m_e)$. MicroBooNE has shown preliminary work measuring electron recoils with kinetic energies as low as 300 keV \cite{Acciarri:2017sjy}. If this can be achieved, it is conceivable that the combined sensitivity of LSND, SBND, MicroBooNE, DUNE and SHiP will provide the leading sensitivity to MCPs in the full range of $5~\MeV \lesssim m_\chi \lesssim 5~\GeV$. 

Further progress may come from new experimental concepts and innovations. Significant progress could come from coupling large underground neutrino detectors with purposely installed new accelerators \cite{Kahn:2014sra,Izaguirre:2015pva}. 
Millicharged particles may also be searched by experiments in {\em disappearance} channels \cite{Lees:2017lec,Banerjee:2016tad,Izaguirre:2014bca,Berlin:2018bsc}, where 
$e^+e^- \to \gamma + \chi + \bar \chi $ and $Z+e^- \to Z+ e^- + \chi + \bar \chi$ production leads to anomalous missing momentum/energy from the $\chi$-pair that pass through a detector without depositing energy. 
Because of the advantageous scaling with $\epsilon$ (second, rather than the fourth power), there are clear prospects on improving bounds on MCPs above the $100$\,MeV energy range. 

In addition, it is also interesting to place a milliQan-type
detector \cite{Haas:2014dda,Ball:2016zrp} downstream of a proton-fixed target along
the Fermilab NuMI/LBNF or CERN SPS beams. The high flux
of MCPs produced on the fixed-target combined with the
dedicated milliQan MCP detector could yield a much more
improved sensitivity reach of MCPs with masses below
5 GeV. This idea is implemented in \cite{Kelly:2018brz} and is named Fermilab MINIcharge-search (FerMINI) experiment.
\\
~

\section{Acknowledgment}
We thank Patrick deNiverville, Ornella Palamara, and Zarko Pavlovic for helping us to understand detector capabilities at Mini- and MicroBooNE. We thank Luca Buonocore, Antonia Di Crescenzo, and Claudia Frugiuele for information about the SHiP detector.
We would also like to thank Matheus Hostert, Dan Hooper, Matthew Klimek, Gordan Krnjaic, Pedro Machado, and Tracy Slatyer for useful discussions. RP and GM are supported by the National Science and Engineering Research Council of Canada (NSERC). YT is supported in part by the U.S. National Science Foundation through grant PHY-1316222. 

This research was supported in part by Perimeter Institute for Theoretical Physics. Research at Perimeter Institute is supported by the Government of Canada through the Department of Innovation, Science and Economic Development and by the Province of Ontario through the Ministry of Research and Innovation.
This manuscript has been authored by Fermi Research Alliance, LLC under Contract No. DE-AC02-07CH11359 with the U.S. Department of Energy, Office of Science, Office of High Energy Physics.


\bibliography{MCP_LSND_21cm}

\begin{thebibliography}{10}
\bibitem{NoteX}
Given \cref{eq:NumberSignalEvents} and $\alpha =
  \Gamma(1+n,b+s_\text{up})/\Gamma(1+n,b)$, any of our limits on $\epsilon$ can
  be rescaled for different choices of backgrounds. This is important for our
  projected sensitivities since detailed modelling of backgrounds will likely
  differ from what is assumed here.
\end{thebibliography}
\bibliographystyle{hunsrt}
\end{document}